\documentclass[12pt]{article}
\usepackage{amssymb}

\def\be{\begin{equation}}
\def\ee{\end{equation}}
\def\bea{\begin{eqnarray}}
\def\eea{\end{eqnarray}}
\def\ba{\begin{array}}
\def\ea{\end{array}}

\def\G{\Gamma}

\def\o{\omega}

\def\np#1{{\sl Nucl.~Phys.~\bf B#1}}
\def\pl#1{{\sl Phys.~Lett.~\bf B#1}}
\def\pr#1{{\sl Phys.~Rev.~\bf D#1}}

\def\cqg#1{{\sl Class.~Quant.~Grav.~\bf #1}}

\catcode`\@=11
\def\@citex[#1]#2{%
\if@filesw \immediate \write \@auxout {\string \citation {#2}}\fi
\@tempcntb\m@ne \let\@h@ld\relax \def\@citea{}%
\@cite{%
  \@for \@citeb:=#2\do {%
    \@ifundefined {b@\@citeb}%
      {\@h@ld\@citea\@tempcntb\m@ne{\bf ?}%
      \@warning {Citation `\@citeb ' on page \thepage \space undefined}}%
      {\@tempcnta\@tempcntb \advance\@tempcnta\@ne%
      \@tempcntb\number\csname b@\@citeb \endcsname \relax%
      \ifnum\@tempcnta=\@tempcntb %Number follows previous--hold on to it
        \ifx\@h@ld\relax%
          \edef \@h@ld{\@citea\csname b@\@citeb\endcsname}%
        \else%
          \edef\@h@ld{\ifmmode{-}\else--\fi\csname b@\@citeb\endcsname}%
        \fi%
      \else%   %  non-successor--dump what's held and do this one
        \@h@ld\@citea\csname b@\@citeb \endcsname%
        \let\@h@ld\relax%
      \fi}%
    \def\@citea{,\penalty\@highpenalty\,}%
  }\@h@ld
}{#1}}

\def\@citeb#1#2{{[#1]\if@tempswa , #2\fi}}
\def\@citeu#1#2{{$^{#1}$\if@tempswa , #2\fi }}
\def\@citep#1#2{{#1\if@tempswa , #2\fi}}

\def\bcites{         % cite with []'s
        \catcode`\@=11
        \let\@cite=\@citeb
        \catcode`\@=12
}

\def\upcites{         % cite with exponents
        \catcode`\@=11
        \let\@cite=\@citeu
        \catcode`\@=12
}

\def\plaincites{      % cite without brackets
        \catcode`\@=11
        \let\@cite=\@citep
        \catcode`\@=12
}

\newcount\hour
\newcount\minute
\newtoks\amorpm
\hour=\time\divide\hour by 60
\minute=\time{\multiply\hour by 60 \global\advance\minute by-\hour}
\edef\standardtime{{\ifnum\hour<12 \global\amorpm={am}%
        \else\global\amorpm={pm}\advance\hour by-12 \fi
        \ifnum\hour=0 \hour=12 \fi
        \number\hour:\ifnum\minute<10 0\fi\number\minute\the\amorpm}}
\edef\militarytime{\number\hour:\ifnum\minute<10 0\fi\number\minute}

\def\draftlabel#1{{\@bsphack\if@filesw {\let\thepage\relax
   \xdef\@gtempa{\write\@auxout{\string
      \newlabel{#1}{{\@currentlabel}{\thepage}}}}}\@gtempa
   \if@nobreak \ifvmode\nobreak\fi\fi\fi\@esphack}
        \gdef\@eqnlabel{#1}}
\def\@eqnlabel{}
\def\@vacuum{}
\def\marginnote#1{}
\def\draftmarginnote#1{\marginpar{\raggedright\scriptsize\tt#1}}
\def\draft{
        \pagestyle{plain}
        \overfullrule=2pt
        \oddsidemargin -.5truein
        \def\@oddhead{\sl \phantom{\today\quad\militarytime} \hfil
        \smash{\Large\sl DRAFT} \hfil \today\quad\militarytime}
        \let\@evenhead\@oddhead
        \let\label=\draftlabel
        \let\marginnote=\draftmarginnote
        \def\ps@empty{\let\@mkboth\@gobbletwo
        \def\@oddfoot{\hfil \smash{\Large\sl DRAFT} \hfil}
        \let\@evenfoot\@oddhead}
        \def\@eqnnum{(\theequation)\rlap{\kern\marginparsep\tt\@eqnlabel}%
        \global\let\@eqnlabel\@vacuum}  }
\title{Low frequency quasi-normal modes of AdS black holes\footnote{Research supported in part by the DoE under grant DE-FG05-91ER40627.}}
\author{George Siopsis\footnote{siopsis@tennessee.edu}\\
\em Department of Physics
and Astronomy, \\
\em The University of Tennessee, Knoxville, \\
\em TN 37996 - 1200, USA.
}
\date{January 2007}
\begin{document}

\maketitle
\vspace{-3.5in}\hfill UTHET-07-0101\vspace{3.5in}

\abstract{We calculate analytically low frequency quasi-normal modes of gravitational perturbations of AdS Schwarzschild black holes in $d$ dimensions.
We arrive at analytic expressions which are in agreement with their counterparts from linearized hydrodynamics in $S^{d-2}\times \mathbb{R}$, in accordance with the AdS/CFT correspondence.
Our results are also in good agreement with results of numerical calculations.
}

\newpage

\section{Introduction}

Quasi-normal modes (QNMs) determine the late-time evolution of black hole perturbations.
They have been extensively studied in asymptotically
Anti-de Sitter (AdS) space-times \cite{qnm} in hopes of shedding some light on
the Anti-de Sitter - conformal field theory (AdS/CFT) correspondence.
Analytic expressions for asymptotic frequencies were derived in~\cite{CNS,NS} (adapting the monodromy argument
proposed in~\cite{MN} and extended to first order in~\cite{SuSi})
for arbitrary dimension.
First-order corrections to these analytic expressions were obtained in \cite{emeis}
in good agreement with numerical results~\cite{CKL}.

At the other end of the spectrum, the low frequency modes can be used to probe the behavior of the gauge theory on the boundary of AdS in the hydrodynamic limit \cite{bibhy}.
This is particularly interesting in connection with heavy ion collisions at RHIC (see, e.g., \cite{bibRHIC} and references therein) and the LHC.
It appears that the quark-gluon plasma (QGP) that forms is strongly coupled raising the possibility that it may possess a gravity dual.
In \cite{bibgub}, the QGP was analyzed in terms of a ``conformal soliton flow'' by mapping the boundary of AdS$_5$, $S^3\times \mathbb{R}$, to the four-dimensional flat Minkowski space by a conformal transformation.
The QNMs were calculated numerically and led to an elliptic flow coefficient and thermalization time that compared well with experimental results.

Here we calculate the low frequency QNMs of AdS Schwarzschild black holes in $d$ dimensions analytically following the procedure of ref.~\cite{bibhy}.
We find agreement with numerical results in four \cite{CKL} and five \cite{bibgub} dimensions.
Our analytic expressions also agree with their counterparts obtained from hydrodynamics on $S^{d-2}\times \mathbb{R}$ \cite{bibgub,bib4d} in accordance with the AdS/CFT correspondence.
We discuss gravitational perturbations using the Master Equation of Ishibashi and Kodama~\cite{IK}, including vector (section~\ref{sect2}), scalar (section~\ref{sect3}) and tensor (section~\ref{sect4}) modes.
Our conclusions are summarized in section~\ref{sect5}.

\section{Vector perturbations}\label{sect2}

We wish to study QNMs of
AdS Schwarzschild black holes whose metric in $d$ dimensions is given by
\be\label{line}
ds^2 = -f(r)dt^2 +\frac{ dr^2 }{ f(r) } +r^2d \Omega_{d-2}^2\ \ ,\ \ \ f(r) = \frac{r^2}{R^2}+1-\frac{2\mu}{r^{d-3}} \;.
\ee
To simplify the notation, we shall set the AdS radius $R=1$.
The radius of the horizon is related to the mass parameter by
\be\label{eqmurH} 2\mu = r_H^{d-1} \left( 1 + \frac{1}{r_H^2} \right) \ee
The radial wave equation for gravitational perturbations in the black-hole
background~(\ref{line}) can be cast into a Schr\"odinger-like form,
\be\label{sch}
  -\frac{d^2\Psi}{dr_*^2}+V[r(r_*)]\Psi =\o^2\Psi \;,
\ee
in terms of the tortoise coordinate defined by
\be\label{tortoise}
  \frac{dr_*}{dr} = \frac{1}{f(r)}\;.
\ee
It is also useful to introduce the distance
\be \bar r_* = r_* (\infty) - r_* (0) = \int_0^\infty \frac{dr}{f(r)} \ee
A short calculation yields \cite{NS}
\be {\bar r}_* = \frac{\pi}{(d-1)r_H} \left( \cot\frac{\pi }{d-1} + i \right) + O(1/r_H^2) \ee
The potential $V$ is determined by the type of perturbation and may be
deduced from the Master Equation  of Ishibashi and Kodama~\cite{IK}.
For vector perturbations, we obtain~\cite{NS}
%\be\label{eqVT} V_{\mathsf{T}} (r) = f(r) \left\{ \frac{\ell (\ell +d-3)}{r^2} + \frac{(d-2)(d-4) f(r)}{4r^2} + \frac{(d-2) f'(r)}{2r} \right\} \ee
\be\label{eqVV} V(r) = V_{\mathsf{V}}(r) \equiv f(r) \left\{ \frac{\ell (\ell +d-3)}{r^2} + \frac{(d-2)(d-4) f(r)}{4r^2} - \frac{r f'''(r)}{2(d-3)} \right\} \ee
%\bea\label{eqVS} V_{\mathsf{S}}(r) &=& \frac{f(r)}{4r^2} \left[ \ell (\ell +d-3) - (d-2) + \frac{(d-1)(d-2)\mu}{r^{d-3}} \right]^{-2} \nonumber\\
%&\times& \Bigg\{ \frac{d(d-1)^2(d-2)^3 \mu^2}{R^2r^{2d-8}}
%- \frac{6(d-1)(d-2)^2(d-4)[\ell (\ell+d-3) - (d-2)] \mu}{R^2r^{d-5}}\nonumber\\
%&& + \frac{(d-4)(d-6)[\ell (\ell+d-3) - (d-2)]^2 r^2}{R^2} +
%\frac{2(d-1)^2(d-2)^4 \mu^3}{r^{3d-9}}\nonumber\\
%&& + \frac{4(d-1)(d-2)(2d^2-11d+18)[\ell (\ell+d-3) - (d-2)]\mu^2}{r^{2d-6}}\nonumber\\
%&& + \frac{(d-1)^2(d-2)^2(d-4)(d-6)\mu^2}{r^{2d-6}}
%- \frac{6(d-2)(d-6)[\ell (\ell+d-3) - (d-2)]^2 \mu}{r^{d-3}}\nonumber\\
%&& - \frac{6(d-1)(d-2)^2(d-4)[\ell (\ell+d-3) - (d-2)] \mu}{r^{d-3}}\nonumber\\
%&& + 4 [\ell (\ell+d-3) - (d-2)]^3 + d(d-2) [\ell (\ell+d-3) - (d-2)]^2 \Bigg\} \eea
Evidently, the potential vanishes at the horizon ($V(r_H) = 0$, since $f(r_H)=0$).
This is the case for all types of perturbation.

%\subsection{Harmonics of order $N\ge 1$}

%Here we discuss gravitational perturbations.
%We extend the procedure of~\cite{CNS,NS} to include first-order
%corrections to analytical expressions for quasi-normal frequencies~\cite{emeis}.
%Our results are in good agreement with numerical results~\cite{CKL}.

The asymptotic form of QNMs was found in \cite{CNS,NS} analytically. Subsequently,
it was shown in \cite{emeis} how the approximation may be improved upon by a perturbative expansion.
One obtains
%We arrive at the first-order expression for quasi-normal frequencies,
\bea\label{eqo1st}
\omega_n &=& \omega_n^{(0)} + \omega_n^{(1)} + \dots \nonumber\\
\omega_n^{(0)} {\bar r}_* &=& \left( n + \frac{d-3}{4} \right) \pi +\frac{1}{2i}\ln 2 \nonumber\\
\omega_n^{(1)} {\bar r}_* &=& \mathcal{A}_\mathsf{V}\frac{e^{-\frac{ i\pi}{2(d-2)}}\cos^4 \frac{ \pi}{2(d-2)}}{2\pi^2 r_H^2} \frac{(\omega_n^{(0)}/r_H)^{-\frac{d-3}{d-2}}}{[2(d-2) (1+1/r_H^2)]^{\frac{1}{d-2}}} \frac{\G(\frac{1}{d-2})\G^4(\frac{d-3}{2(d-2)})(d-3)}{(d-1)}
\eea
where
\be \mathcal{A}_\mathsf{V} = \frac{\ell(\ell+d-3)}{d-2} + \frac{d^2-8d+13}{2(2d-15)} \ee
%\[ {\bar r}_* = \int_0^\infty \frac{dr}{f(r)} = \frac{\pi}{(d-1)r_H} \left( \cot\frac{\pi }{d-1} + i \right) + O(1/r_H^2) \]
%where we took the limit of interest $j\to 0,2$ wherever it was unambiguous, in
%order to simplify the notation.
Thus, the first-order correction is $O(n^{-\frac{d-3}{d-2}})$.
It should also be noted that the first-order correction is suppressed by a factor of $1/r_H^2$.
So for large black holes, the zeroth-order contribution provides a good approximation to all modes, not just the high overtones.
Moreover, the zeroth-order term is independent of both the angular momentum quantum number $\ell$ and the type of perturbation. This leads to a mild dependence on $\ell$ and the type of perturbation of all QNMs for large black holes ($r_H \gtrsim 1$).
Finally, since $\bar r_* \sim 1/r_H$, the frequencies are proportional to the radius of the horizon.

In particular, for $d=5$, the QNMs are given by
\be\label{eq1} \frac{\omega_n}{r_H} = \left(2n+1 +i \frac{\ln 2}{\pi} \right)
(1 - i) + O(1/r_H^2) \ee
Numerically, we find
\be\label{eq2} \frac{\omega_1}{r_H} \approx \left( 3+ i\frac{\ln 2}{\pi} \right) (1-i) = 3.221-2.779 i \ \ , \ \ \ \  \frac{\Delta\omega_n}{r_H} \equiv \frac{\omega_{n+1} - \omega_n}{r_H} \approx 2(1-i) \ee
These results are independent of the type of perturbation and the quantum number $\ell$.
They agree well with numerical results~\cite{bibgub}.

% the values listed in Table~\ref{tab1}.
%They are independent of the type of perturbation as well as the quantum number $n$.
%These are $O(n^{-2/3})$ effects which have already been calculated in \cite{emeis}.

%Notice also that for large $r_H$, (\ref{eq1}) reduces to
%\be \omega_1/r_H \approx \left( 3+ i\frac{\ln 2}{\pi} \right) (1-i) = 3.221-2.779 i \ \ , \ \ \ \  \Delta\omega/r_H \approx 2(1-i) \ee

%\begin{table}[h]
%\begin{center}
%\begin{tabular}{|l|l|l|l|l|} \hline
%$r_H$ & $\omega_1/r_H$ & $\omega_2/r_H$ & $\omega_3/r_H$ & $\omega_4/r_H$
%\\\hline
%$6$ & $3.262-2.776 i$ & $5.290-4.776 i$ & $7.317-6.776 i$ & $9.345-8.776i$
%\\\hline
%$13$ & $3.229-2.779 i$ & $5.235-4.779 i$ & $7.241-6.779 i$ & $9.247-8.779i$
%\\\hline
%$20$ & $3.224-2.779 i$ & $5.227-4.779 i$ & $7.229-6.779 i$ & $9.232-8.779i$
%\\\hline
%\end{tabular}
%\end{center}
%\caption{Analytical QNMs}\label{tab1}
%\end{table}

%First-order $O(n^{-2/3})$ corrections are small ($\sim r_H^{-2}$).

%\subsection{Zero modes ($N=0$)}

%I'll do perturbations in the limit of large horizon radius.

%\subsubsection*{Vector perturbations}

Even though the above expressions include high as well as low frequencies,
they do not include the lowest overtones.
The latter may be obtained using the method of \cite{bibhy} and they correspond to the hydrodynamic behavior of the gauge theory fluid.

To obtain analytic expressions for the lowest overtones, it is convenient to introduce the coordinate
\be\label{eqru} u = \left( \frac{r_H}{r} \right)^{d-3} \ee
The wave equation (\ref{sch}) becomes
\be\label{eq13} -(d-3)^2 u^{\frac{d-4}{d-3}}\hat f(u) \left( u^{\frac{d-4}{d-3}}\hat f(u) \Psi' \right)' +\hat V_{\mathsf{V}} (u)\Psi = \hat\omega^2 \Psi  \ \ , \ \ \ \
\hat\omega = \frac{\omega}{r_H}\ee
where prime denotes differentiation with respect to $u$ and we have defined
\be\label{eq14} \hat f(u) \equiv \frac{f(r)}{r^2} = 1- u^{\frac{2}{d-3}} \left( u- \frac{1 - u}{r_H^2} \right) \ee
\be\label{eq15} \hat V_{\mathsf{V}} (u) \equiv \frac{V_{\mathsf{V}}}{r_H^2} = \hat f(u) \left\{ \hat L^2 + \frac{(d-2)(d-4)}{4} u^{-\frac{2}{d-3}}\hat f(u) - \frac{(d-1)(d-2)\left( 1+ \frac{1}{r_H^2} \right)}{2} u\right\} \ee
where
\be\label{eq16} \hat L^2 = \frac{\ell (\ell +d-3)}{r_H^2} \ee
%For simplicity,
Let us first consider the large black hole limit $r_H \to\infty$ keeping $\hat\omega$ and $\hat L$ fixed (small).
%The wave equation simplifies to
%\[ - (d-3)^2 (u^{\frac{2d-8}{d-3}} -u^3)\Psi'' - (d-3) [ (d-4)u^{\frac{d-5}{d-3}}-(2d-5)u^2]\Psi' \]
%\be + \left\{ \hat L^2 + \frac{(d-2)(d-4)}{4}u^{-\frac{2}{d-3}} - \frac{3(d-2)^2}{4} u - \frac{\hat\omega^2}{1-u^{\frac{d-1}{d-3}}} \right\} \Psi = 0 
%\ee
Factoring out the behavior at the horizon ($u=1$)
% (horizon) and $u=0$ (boundary),
\be \Psi = (1-u)^{-i \frac{\hat\omega}{d-1}} F(u) \ee
the wave equation simplifies to
%we obtain
\be\label{sch2} \mathcal{A} F'' + \mathcal{B}_{\hat\omega} F' + \mathcal{C}_{\hat\omega , \hat L} F = 0 \ee
where
\bea \mathcal{A} &=& - (d-3)^2 u^{\frac{2d-8}{d-3}} (1-u^{\frac{d-1}{d-3}}) \nonumber\\
\mathcal{B}_{\hat\omega} &=& - (d-3) [ d-4-(2d-5)u^{\frac{d-1}{d-3}}]u^{\frac{d-5}{d-3}} - 2(d-3)^2 \frac{i\hat\omega}{d-1}\frac{u^{\frac{2d-8}{d-3}} (1-u^{\frac{d-1}{d-3}})}{1-u} \nonumber\\
\mathcal{C}_{\hat\omega , \hat L} &=& \hat L^2 + \frac{(d-2)[d-4-3(d-2)u^{\frac{d-1}{d-3}}]}{4}u^{-\frac{2}{d-3}} \nonumber\\
& & - \frac{\hat\omega^2}{1-u^{\frac{d-1}{d-3}}} + (d-3)^2 \frac{\hat\omega^2}{(d-1)^2}\frac{u^{\frac{2d-8}{d-3}} (1-u^{\frac{d-1}{d-3}})}{(1-u)^2}\nonumber\\
& &
- (d-3)\frac{i\hat\omega}{d-1} \frac{[ d-4-(2d-5)u^{\frac{d-1}{d-3}}]u^{\frac{d-5}{d-3}} }{1-u} - (d-3)^2 \frac{i\hat\omega}{d-1}\frac{u^{\frac{2d-8}{d-3}} (1-u^{\frac{d-1}{d-3}})}{(1-u)^2}\nonumber\eea
For small $\hat\omega$, $\hat L$, eq.~(\ref{sch2}) may be solved perturbatively by writing it as
\be (\mathcal{H}_0 + \mathcal{H}_1) F = 0 \ee
where
\bea\label{eqH0} \mathcal{H}_0 F &\equiv& \mathcal{A} F'' + \mathcal{B}_0 F' + \mathcal{C}_{0 , 0} F \nonumber\\
\mathcal{H}_1 F &\equiv& (\mathcal{B}_{\hat\omega} - \mathcal{B}_0) F' + (\mathcal{C}_{\hat\omega , \hat L} - \mathcal{C}_{0 , 0}) F \eea
Expanding the wavefunction perturbatively,
\be F = F_0 + F_1 + \dots \ee
at zeroth order we have
\be\label{eq22} \mathcal{H}_0 F_0 = 0 \ee
whose acceptable solution is
\be\label{eq23} F_0 = u^{\frac{d-2}{2(d-3)}} \ee
It is regular at both the horizon ($u=1$) and the boundary ($u=0$, or $\Psi \sim r^{-\frac{d-2}{2}}\to 0$ as $r\to\infty$).
The Wronskian (up to an arbitrary multiplicative constant) is
\be \mathcal{W} = \frac{1}{u^{\frac{d-4}{d-3}} (1-u^{\frac{d-1}{d-3}})} \ee
Another linearly independent solution is
\be\label{eqchF0} \check F_0 = F_0\int \frac{\mathcal{W}}{F_0^2} \ee
It is unacceptable because it diverges at both the horizon ($\check F_0 \sim \ln (1-u)$ for $u\approx 1$) and the boundary ($\check F_0 \sim u^{-\frac{d-4}{2(d-3)}}$ for $u\approx 0$, or $\Psi \sim r^{\frac{d-4}{2}} \to\infty$ as $r\to\infty$).
It may be expressed in terms of hypergeometric functions but its explicit form is not needed for our purposes (first-order perturbation theory).

At first order we have
\be \mathcal{H}_0 F_1 =- \mathcal{H}_1 F_0 \ee
whose solution may be written as
\be\label{eqF1} F_1 = F_0\int \frac{\mathcal{W}}{F_0^2} \int \frac{F_0\mathcal{H}_1 F_0}{\mathcal{A}\mathcal{W}} \ee
The limits of the inner integral may be adjusted at will because this amounts to adding an arbitrary amount of the unacceptable solution (\ref{eqchF0}).
To ensure regularity at the horizon, we should choose one of the limits at $u=1$ (the integrand is regular at the horizon, by design).
Then the behavior of the wavefunction (\ref{eqF1}) at the boundary ($u=0$) is given by
\be F_1 = \check F_0 \int_0^1 \frac{F_0\mathcal{H}_1 F_0}{\mathcal{A}\mathcal{W}} + \dots \ee
where we omitted regular terms.
The coefficient of the singularity ought to vanish,
\be\label{eq29} \int_0^1 \frac{F_0 \mathcal{H}_1 F_0}{\mathcal{A}\mathcal{W}} = 0 \ee
which imposes a constraint on the parameters (dispersion relation) of the form
\be\label{eqdisp} \mathbf{a}_0 \hat L^2 -i \mathbf{a}_1 \hat\omega - \mathbf{a}_2 \hat\omega^2 = 0 \ee
After some algebra, we arrive at explicit expressions for the coefficients,
\be\label{eqcoef} \mathbf{a}_0 = \frac{d-3}{d-1} \ \ , \ \ \ \
\mathbf{a}_1 = d-3 \ee
The coefficient $\mathbf{a}_2$ may also be found explicitly for each dimension $d$, but it cannot be written as a function of $d$ in closed form.
However, it does not contribute to the dispersion relation at lowest order.
E.g., for $d=4,5$, we obtain, respectively
\be\label{eqa2} \mathbf{a}_2 = \frac{65}{108} -\frac{1}{3}\ln 3 \ \ , \ \ \ \
\frac{5}{6}-\frac{1}{2}\ln 2 \ee
%\[ \mathbf{a}_0 = \int_0^1 u^{\frac{2}{d-3}} = \frac{d-3}{d-1}\]
%\[ \mathbf{a}_1 = -(d-3)\int_0^1 u^{\frac{2}{d-3}} \left[ (d-2) \frac{u^{\frac{2d-8}{d-3}} -u^3}{u(1-u)} + \frac{ (d-4)u^{\frac{d-5}{d-3}}-(2d-5)u^2}{1-u}
% + (d-3) \frac{u^{\frac{2d-8}{d-3}} -u^3}{(1-u)^2}\right] \]
%\[ = - (d-1)(d-3) \]
%\[ \mathbf{a}_2 = \int_0^1 u^{\frac{2}{d-3}}\left[ (d-3)^2 \frac{u^{\frac{2d-8}{d-3}} -u^3}{(1-u)^2} - \frac{(d-1)^2}{1-u^{\frac{d-1}{d-3}}} \right] \]
%\[ - (1-u^{d-1}) F'' + \left\{ (d-1) u^{d-2} -2 i \frac{\hat\omega}{d-1} \frac{1-u^{d-1}}{1-u} \right\} F' \]
%For $d=5$, $\mathbf{a}_2 = \frac{5}{6}-\frac{1}{2}\ln 2$.
%For $d=4$, $\mathbf{a}_2 = \frac{65}{108} -\frac{1}{3}\ln 3$.
The quadratic in $\hat\omega$ equation (\ref{eqdisp}) has two solutions,
\be \hat\omega_0 \approx -i\frac{\hat L^2}{d-1} \ \ , \ \ \ \  \hat\omega_1 \approx -i \frac{d-3}{\mathbf{a}_2} + i\frac{\hat L^2}{d-1} \ee
where we omitted terms of higher order in $\hat L^2$.
In terms of the frequency $\omega$ and the quantum number $\ell$, they may be written respectively as
\be\label{eq34} \omega_0 \approx -i\frac{\ell(\ell+d-3)}{(d-1)r_H} \ \ , \ \ \ \  \frac{\omega_1}{r_H} \approx -i \frac{d-3}{\mathbf{a}_2} + i\frac{\ell(\ell+d-3)}{(d-1)r_H^2} \ee
The smaller of the two, $\omega_0$, is inversely proportional to the radius of the horizon and is not included in the spectrum (\ref{eqo1st}) obtained earlier by different means \cite{CNS,NS,emeis}.
The other solution, $\omega_1$ is a crude estimate of the first overtone in the spectrum (\ref{eqo1st}).
Numerically, for $d=5$, we obtain using (\ref{eqa2})
\be\label{eqod5} \frac{\omega_1}{r_H} \approx -4.109 i + i \frac{\ell (\ell +2)}{4r_H^2} \ee
to be compared with the numerical value (\ref{eq2}).

It should be noted that the crude estimate (\ref{eqod5}) already exhibits two important features: $\omega_1$ is proportional to $r_H$ and the dependence on $\ell$ is of order $1/r_H^2$, as expected from (\ref{eqo1st}).
The approximation may be improved by including higher-order terms in the perturbative expansion.
Inclusion of higher orders also increases the degree of the polynomial in the dispersion relation (\ref{eqdisp}) whose roots then yield approximate values of more QNMs. Thus, this method reproduces the spectrum (\ref{eqo1st}) albeit not in an efficient way.

The above discussion may be extended to black holes of finite size in a straightforward manner by treating finite-size effects as a perturbation
(assuming $1/r_H$ is small).
Thus at first order, we need to replace $\mathcal{H}_1$ (eq.~(\ref{eqH0})) by
\be\label{eqH1} \mathcal{H}_1' = \mathcal{H}_1 + \frac{1}{r_H^2} \mathcal{H}_H \ee
where
%For finite $r_+$, we may add
\be \mathcal{H}_H F \equiv \mathcal{A}_H F'' + \mathcal{B}_H F' + \mathcal{C}_H F \ee
The coefficients may be easily deduced by collecting $O(1/r_H^2)$ terms in the exact wave equation given by (\ref{eq13}), (\ref{eq14}) and (\ref{eq15}).
We obtain
\bea \mathcal{A}_H &=& -2(d-3)^2 u^2(1-u) \nonumber\\
\mathcal{B}_H &=& -(d-3) u\left[ (d-3)(2-3u) - (d-1) \frac{1-u}{1-u^{\frac{d-1}{d-3}}} u^{\frac{d-1}{d-3}} \right] \nonumber\\
\mathcal{C}_H &=& \frac{d-2}{2} \left[ d-4-(2d-5)u - (d-1) \frac{1-u}{1-u^{\frac{d-1}{d-3}}} u^{\frac{d-1}{d-3}} \right] \eea
Interestingly, the zeroth order wavefunction $F_0$ (eq.~(\ref{eq23})) is an eigenfunction of $\mathcal{H}_H$,
\be \mathcal{H}_H F_0 = -(d-2) F_0 \ee
therefore, the first-order finite-size effect is a simple shift of the angular momentum (eq.~(\ref{eq16}))
\be \hat L^2 \to \hat L^2 - \frac{d-2}{r_H^2} \ee
Consequently, the QNMs of lowest frequency (\ref{eq34}) are modified to
\be\label{eqo0} \omega_0 = - i \frac{\ell(\ell+d-3)-(d-2)}{(d-1)r_H} + O(1/r_H^2) \ee
For $d=4, 5$, we have respectively,
\be \omega_0 = - i \frac{(\ell-1)(\ell+2)}{3r_H} \ \ , \ \ \ \  - i \frac{(\ell+1)^2-4}{4r_H} \ee
in agreement with numerical results (\cite{CKL} and \cite{bibgub}, respectively).

Next we discuss the role the lowest frequency mode plays in the AdS/CFT correspondence.
The dual to the AdS Schwarzschild black hole is a gauge theory fluid on the boundary of AdS ($S^{d-2} \times \mathbb{R}$).
To find the dual of vector perturbations, one ought to consider the fluid dynamics ansatz
\be\label{eqansv} u_i = \mathcal{K} e^{-i\Omega \tau} \mathbb{V}_i \ee
where $u_i$ is the (small) velocity of a point in the fluid and $\mathbb{V}_i$ is a vector harmonic on $S^{d-2}$.
Demanding that this ansatz satisfy the standard equations of linearized hydrodynamics, one arrives at a constraint on the frequency of the perturbation $\Omega$ which yields \cite{bibgub,bib4d}
\be \Omega = -i \frac{\ell(\ell+d-3)-(d-2)}{(d-1)r_H} + O(1/r_H^2) \ee
in perfect agreement with its dual counterpart (\ref{eqo0}).

\section{Scalar perturbations}\label{sect3}

Scalar perturbations are also governed by a Schr\"odinger-like wave equation (\ref{sch})
with the potential given by \cite{IK}
\bea  V_{\mathsf{S}}(r) &=& \frac{f(r)}{4r^2} \left[ \ell (\ell +d-3) - (d-2) + \frac{(d-1)(d-2)\mu}{r^{d-3}} \right]^{-2} \nonumber\\
&\times& \Bigg\{ \frac{d(d-1)^2(d-2)^3 \mu^2}{r^{2d-8}}
- \frac{6(d-1)(d-2)^2(d-4)[\ell (\ell+d-3) - (d-2)] \mu}{r^{d-5}}\nonumber\\
&& + (d-4)(d-6)[\ell (\ell+d-3) - (d-2)]^2 r^2 +
\frac{2(d-1)^2(d-2)^4 \mu^3}{r^{3d-9}}\nonumber\\
&& + \frac{4(d-1)(d-2)(2d^2-11d+18)[\ell (\ell+d-3) - (d-2)]\mu^2}{r^{2d-6}}\nonumber\\
&& + \frac{(d-1)^2(d-2)^2(d-4)(d-6)\mu^2}{r^{2d-6}}
- \frac{6(d-2)(d-6)[\ell (\ell+d-3) - (d-2)]^2 \mu}{r^{d-3}}\nonumber\\
&& - \frac{6(d-1)(d-2)^2(d-4)[\ell (\ell+d-3) - (d-2)] \mu}{r^{d-3}}\nonumber\\
&& + 4 [\ell (\ell+d-3) - (d-2)]^3 + d(d-2) [\ell (\ell+d-3) - (d-2)]^2 \Bigg\} \eea
QNMs are asymptotically the same as the QNMs (\ref{eqo1st}) of vector perturbations \cite{CNS,NS}.
They differ at first order in the perturbative expansion. For scalar modes we obtain \cite{emeis}
\be\label{eqo1sc} \omega_n^{(1)} {\bar r}_* = \mathcal{A}_\mathsf{S}\frac{e^{-\frac{ i\pi}{2(d-2)}}\cos^4 \frac{ \pi}{2(d-2)}}{2\pi^2 r_H^2} \frac{(\omega_n^{(0)}/r_H)^{-\frac{d-3}{d-2}}}{[2(d-2) (1+1/r_H^2)]^{\frac{1}{d-2}}} \G(\frac{1}{d-2})\G^4(\frac{d-3}{2(d-2)})
\ee
where
\be \mathcal{A}_\mathsf{S} = \frac{(d^2-7d+14)[\ell(\ell+d-3)-(d-2)]}{(d-1)(d-2)^2} + \frac{2d^3-24d^2+94d-116}{4(2d-5)(d-2)} \ee
Again, we see the $\ell$-dependence entering at $O(1/r_H^2)$, as for vector perturbations.
Also, the above set of frequencies does not exhaust the spectrum as it leaves out the lowest frequency mode.
To find it, we work as before. Changing variables to (\ref{eqru}), we may write the wave equation as in (\ref{eq13}) with $\hat V_{\mathsf{V}}$ replaced by
\bea  \hat V_{\mathsf{S}}(u) &=& \frac{\hat f(u)}{4} \left[ \hat m + \left( 1 + \frac{1}{r_H^2} \right) u\right]^{-2}  \nonumber\\
&\times& \Bigg\{ d(d-2) \left( 1+ \frac{1}{r_H^2} \right)^2 u^{\frac{2d-8}{d-3}}
- 6(d-2)(d-4)\hat m \left( 1+ \frac{1}{r_H^2} \right) u^{\frac{d-5}{d-3}}\nonumber\\
&& + (d-4)(d-6)\hat m^2u^{-\frac{2}{d-3}} +
(d-2)^2 \left( 1+ \frac{1}{r_H^2} \right)^3 u^3 \nonumber\\
&& + 2(2d^2-11d+18)\hat m \left( 1+ \frac{1}{r_H^2} \right)^2 u^2\nonumber\\
&& + \frac{(d-4)(d-6)\left( 1+\frac{1}{r_H^2} \right)^2}{r_H^2} u^2
- 3(d-2)(d-6)\hat m^2 \left( 1+\frac{1}{r_H^2} \right) u\nonumber\\
&& - \frac{6(d-2)(d-4)\hat m\left( 1+\frac{1}{r_H^2} \right)}{r_H^2} u + 2 (d-1)(d-2)\hat m^3 + d(d-2) \frac{\hat m^2}{r_H^2} \Bigg\} \eea
where
\be\label{eqm} \hat m = 2\frac{\ell (\ell+d-3) - (d-2)}{(d-1)(d-2)r_H^2} = \frac{2(\ell + d-2)(\ell -1)}{(d-1)(d-2)r_H^2} \ee
In the large black hole limit $r_H\to \infty$ with $\hat m$ fixed, the potential simplifies to
\bea  \hat V_{\mathsf{S}}^{(0)}(u) &=& \frac{1-u^{\frac{d-1}{d-3}}}{4( \hat m + u)^2}
\Bigg\{ d(d-2) u^{\frac{2d-8}{d-3}}
- 6(d-2)(d-4)\hat m u^{\frac{d-5}{d-3}}\nonumber\\
&& + (d-4)(d-6)\hat m^2u^{-\frac{2}{d-3}} +
(d-2)^2 u^3\nonumber\\
&&
+ 2(2d^2-11d+18)\hat m u^2
- 3(d-2)(d-6)\hat m^2  u
+ 2 (d-1)(d-2)\hat m^3  \Bigg\} \eea
In addition to the singularities at the end points ($u=0,1$), the scalar wave equation
has an additional singularity due to the double pole of the scalar potential
at $u = -\hat m$.
It is desirable to factor out the behavior not only at the horizon, but also at the boundary and the pole of the scalar potential.
We therefore define
\be\label{eqPsiF} \Psi = (1-u)^{-i\frac{\hat\omega}{d-1}} \frac{u^{\frac{d-4}{2(d-3)}}}{\hat m + u} F(u) \ee
In terms of $F(u)$, the wave equation for scalar perturbations in the large black hole limit reads
%Factoring out the bahavior at the horizon, we obtain
\be\label{eqwsc} \mathcal{A} F'' + \mathcal{B}_{\hat\omega} F' + \mathcal{C}_{\hat\omega} F = 0 \ee
where
\bea \mathcal{A} &=& - (d-3)^2 u^{\frac{2d-8}{d-3}} (1-u^{\frac{d-1}{d-3}}) \nonumber\\
\mathcal{B}_{\hat\omega} &=& - (d-3) u^{\frac{2d-8}{d-3}} (1-u^{\frac{d-1}{d-3}}) \left[ \frac{d-4}{u} -\frac{2(d-3)}{\hat m + u} \right] \nonumber\\
&& - (d-3) [ d-4-(2d-5)u^{\frac{d-1}{d-3}}]u^{\frac{d-5}{d-3}} - 2(d-3)^2 \frac{i\hat\omega}{d-1}\frac{u^{\frac{2d-8}{d-3}} (1-u^{\frac{d-1}{d-3}})}{1-u} \nonumber\\
\mathcal{C}_{\hat\omega} &=&  - u^{\frac{2d-8}{d-3}} (1-u^{\frac{d-1}{d-3}}) \left[ -\frac{(d-2)(d-4)}{4 u^2} - \frac{(d-3)(d-4)}{u(\hat m + u)} + \frac{2(d-3)^2}{(\hat m + u)^2} \right] \nonumber\\
&& - \left[ \left\{ d-4-(2d-5)u^{\frac{d-1}{d-3}} \right\} u^{\frac{d-5}{d-3}} + 2(d-3) \frac{i\hat\omega}{d-1}\frac{u^{\frac{2d-8}{d-3}} (1-u^{\frac{d-1}{d-3}})}{1-u}\right]\left[ \frac{d-4}{2u} - \frac{d-3}{\hat m + u} \right] \nonumber\\
&& - (d-3)\frac{i\hat\omega}{d-1} \frac{[ d-4-(2d-5)u^{\frac{d-1}{d-3}}]u^{\frac{d-5}{d-3}} }{1-u} - (d-3)^2 \frac{i\hat\omega}{d-1}\frac{u^{\frac{2d-8}{d-3}} (1-u^{\frac{d-1}{d-3}})}{(1-u)^2}\nonumber\\
&& + \frac{\hat V_{\mathsf{S}}^{(0)}(u)-\hat\omega^2}{1-u^{\frac{d-1}{d-3}}} + (d-3)^2 \frac{\hat\omega^2}{(d-1)^2}\frac{u^{\frac{2d-8}{d-3}} (1-u^{\frac{d-1}{d-3}})}{(1-u)^2}\nonumber\eea
To calculate the QNMs perturbatively, we define the zeroth-order wave equation
as in (\ref{eq22}) with
\be\label{eq0sc} \mathcal{H}_0 F \equiv \mathcal{A} F'' + \mathcal{B}_0 F' \ee
The acceptable zeroth-order solution is (arbitrarily normalized)
\be\label{eqF0sc} F_0(u) = 1 \ee
which is plainly regular at all singular points ($u=0,1, -\hat m$).
On account of (\ref{eqPsiF}), it corresponds to a wavefunction vanishing at the boundary ($\Psi \sim r^{-\frac{d-4}{2}}$ as $r\to\infty$, using (\ref{eqru})).
The Wronskian is
\be\label{eqWsc} \mathcal{W} = \frac{\left( \hat m + u\right)^2 }{u^{\frac{2d-8}{d-3}} (1-u^{\frac{d-1}{d-3}})} \ee
The unacceptable solution is
\be\label{eqF0scu} \check F_0 = \int \mathcal{W} \ee
and can be written in terms of hypergeometric functions.
It has a singularity at the boundary, $\check F_0 \sim u^{-\frac{d-5}{d-3}}$ for $u\approx 0$ (or $\Psi \sim r^{\frac{d-6}{2}}\to\infty$ as $r\to\infty$ for $d\ge 6$; for $d=5$, the acceptable wavefunction behaves as $r^{-1/2}$ whereas the unacceptable wavefunction behaves as $r^{-1/2} \ln r$; for $d=4$ the roles of $F_0$ and $\check F_0$ are reversed, however our results are still valid).
$\check F_0$ is also singular (logarithmically) at the horizon ($u=1$).

At first order in perturbation theory, working as in the case of vector modes,
we arrive at the constraint
\be\label{eqcnssc} \int_0^1 \frac{\hat\mathcal{C}_{\hat\omega}}{\mathcal{A}\mathcal{W}} = 0 \ee
where we used (\ref{eq29}), (\ref{eqF0sc}) and
\be \mathcal{H}_1 F_0 \equiv (\mathcal{B}_{\hat\omega} - \mathcal{B}_0) F_0'
+ \mathcal{C}_{\hat\omega} F_0 = \mathcal{C}_{\hat\omega} \ee
The first-order constraint (\ref{eqcnssc}) may be written as a dispersion relation
\be\label{eqcnssc2} \mathbf{a}_0 - \mathbf{a}_1 i\hat\omega - \mathbf{a}_2 \hat\omega^2 = 0 \ee
After some algebra, we obtain
\be \mathbf{a}_0 = \frac{d-1}{2} \ \frac{ 1+ (d-2)\hat m}{(1+ \hat m )^2} \ \ ,
\ \ \ \ \mathbf{a}_1 = \frac{d-3}{(1+ \hat m )^2} \ \ ,
\ \ \ \ \mathbf{a}_2 = \frac{1}{\hat m} \left\{ 1 + O(\hat m) \right\} \ee
For small $\hat m$, the quadratic equation (\ref{eqcnssc2}) yields the solutions
\be\label{eqosc1} \hat\omega_0^\pm \approx - i\frac{d-3}{2} \ \hat m \pm \sqrt{ \frac{d-1}{2} \ \hat m} \ee
The two solutions are related to each other by $\hat\omega_0^+ = -\hat\omega_0^{-*}$, which is a general symmetry of the spectrum.
Neither solution is included in the spectrum obtained from asymptotic QNMs.
To obtain approximations to those modes, we need to consider higher orders in perturbation theory.
Notice also that unlike vector modes, these scalar lowest frequency modes have finite real part.
Using (\ref{eqm}), we may express the frequencies in terms of the quantum number $\ell$ as
\be\label{eqosc1a} \omega_0^\pm \approx - i(d-3) \ \frac{\ell (\ell+d-3)-(d-2)}{(d-1)(d-2) r_H} \pm \sqrt{ \frac{\ell (\ell+d-3)-(d-2)}{d-2}} \ee
Thus the real part is independent of $r_H$ whereas the imaginary part is inversely proportional to $r_H$ (for $r_H\gtrsim 1$).

Finite size effects may be added perturbatively.
At first order, they amount to a shift of the coefficient $\mathbf{a}_0$ in the dispersion relation (\ref{eqcnssc2}),
\be \mathbf{a}_0 \to \mathbf{a}_0 + \frac{1}{r_H^2} \mathbf{a}_H \ee
Working as in the vector case, after some tedious but straightforward algebra, we obtain
\be \mathbf{a}_H = \frac{1}{\hat m} \left\{ 1 + O(\hat m) \right\} \ee
The modified dispersion relation yields the modes
\be\label{eqosc2} \hat\omega_0^\pm \approx - i\frac{d-3}{2} \ \hat m \pm \sqrt{ \frac{d-1}{2} \ \hat m +1} \ee
correcting (\ref{eqosc1}).
Explicitly,
\be\label{eqosc2a} \omega_0^\pm \approx - i(d-3) \ \frac{\ell (\ell+d-3)-(d-2)}{(d-1)(d-2) r_H} \pm \sqrt{ \frac{\ell (\ell+d-3)}{d-2}} \ee
correcting (\ref{eqosc1a}).
This is in agreement with numerical results \cite{bibgub,bib4d}.

Turning to the AdS/CFT correspondence, we ought to purturb the gauge theory fluid on the boundary of AdS ($S^{d-2} \times \mathbb{R}$) using the ansatz
\be\label{eqanss} u_i = \mathcal{K} e^{-i\Omega \tau} \nabla_i \mathbb{S} \ \ , \ \ \ \
\delta p = \mathcal{K}' e^{-i\Omega \tau} \mathbb{S} \ee
where $u_i$ is the (small) velocity of a point in the fluid, $\delta p$ is the pressure perturbation and $\mathbb{S}_i$ is a scalar harmonic on $S^{d-2}$.
Demanding that this ansatz satisfy the standard equations of linearized hydrodynamics, one arrives at an expression for the frequency of the perturbation $\Omega$ \cite{bibgub,bib4d} which is in perfect agreement with our analytic result (\ref{eqosc2a}).

\section{Tensor perturbations}\label{sect4}

Finally we discuss tensor perturbations. In this case the potential is given by
\cite{IK}
\be\label{eqVT2} V_{\mathsf{T}} (r) = f(r) \left\{ \frac{\ell (\ell +d-3)}{r^2} + \frac{(d-2)(d-4) f(r)}{4r^2} + \frac{(d-2) f'(r)}{2r} \right\} \ee
It leads to the same asymptotic form of QNMs as in the other two cases (vector and scalar) \cite{CNS,NS}.
The first-order correction to the asymptotic expression is \cite{emeis}
\be\label{eqo1t} \omega_n^{(1)} {\bar r}_* = \mathcal{A}_\mathsf{T}\frac{e^{-\frac{ i\pi}{2(d-2)}}\cos^4 \frac{ \pi}{2(d-2)}}{2\pi^2 r_H^2} \frac{(\omega_n^{(0)}/r_H)^{-\frac{d-3}{d-2}}}{[2(d-2) (1+1/r_H^2)]^{\frac{1}{d-2}}} \G(\frac{1}{d-2})\G^4(\frac{d-3}{2(d-2)})
\ee
where
\be \mathcal{A}_\mathsf{T} = \frac{\ell(\ell+d-3)}{d-2} + \frac{(d-3)^2}{2(2d-5)} \ee
to be compared with the vector (\ref{eqo1st}) and scalar (\ref{eqo1sc}) cases.
Once again we obtain a mild $O(1/r_H^2)$ dependence on the quantum number $\ell$.

Unlike the other two cases, this constitutes the entire spectrum.
To see this, let us change variables to (\ref{eqru}) and go to the large black hole limit.
The wave equation simplifies to
\bea\label{eqwavt} - (d-3)^2 (u^{\frac{2d-8}{d-3}} -u^3)\Psi'' - (d-3) [ (d-4)u^{\frac{d-5}{d-3}}-(2d-5)u^2]\Psi' && \nonumber\\
 + \left\{ \hat L^2 + \frac{d(d-2)}{4}u^{-\frac{2}{d-3}} + \frac{(d-2)^2}{4} u - \frac{\hat\omega^2}{1-u^{\frac{d-1}{d-3}}} \right\} \Psi &=& 0 
\eea
The zeroth order equation is obtained from (\ref{eqwavt}) by setting $\hat L$ and $\hat\omega$ to zero.
The resulting equation may be solved exactly.
Two linearly independent solutions are ($\Psi = F_0$ at zeroth order)
\be F_0(u) = u^{\frac{d-2}{2(d-3)}} \ \ , \ \ \ \ \check F_0(u) = u^{-\frac{d-2}{2(d-3)}} \ln \left( 1-u^{\frac{d-1}{d-3}} \right) \ee
Neither behaves nicely at both ends ($u=0,1$), therefore both are unacceptable.
It is not possible to build a perturbation theory to calculate small frequencies.

This negative result is in agreement with numerical results and is also in accordance with the AdS/CFT correspondence \cite{bibgub}.
Indeed, there is no ansatz that can be built from tensor spherical harmonics $\mathbb{T}_{ij}$ (similar to the vector (\ref{eqansv}) and scalar (\ref{eqanss}) cases) satisfying the linearized hydrodynamic equations because of the conservation and tracelessness properties of $\mathbb{T}_{ij}$ \cite{bibgub}.

\section{Conclusion}\label{sect5}

We calculated analytically low frequency QNMs of gravitational perturbations of AdS Schwarzschild black holes in arbitrary dimension using the Master Equation of Ishibashi and Kodama \cite{IK}.
We noted that low frequency modes are well approximated by asymptotic expressions for large black holes (with radius of horizon $r_H \gtrsim 1$ in units such that the AdS radius $R=1$) \cite{CNS,NS}.
These expressions ($\omega_n^{(0)}$ in eq.~(\ref{eqo1st})) are independent of the type of perturbation and the angular momentum quantum number $\ell$.
The dependence on $\ell$ enters at first-order perturbation theory and is $O(1/r_H^2)$ \cite{emeis}.
The first-order contribution, $\omega_n^{(1)}$, for vector, scalar and tensor perturbations is given by eqs.~(\ref{eqo1st}), (\ref{eqo1sc}) and (\ref{eqo1t}), respectively.

Asymptotic expressions do not in general yield the entire spectrum. To find the lowest frequency mode, we applied the method in ref.~\cite{bibhy}.
We also included the effects of a finite size black hole.
We arrived at explicit analytic expressions in the case of vector (eq.~(\ref{eqo0})) and scalar (eq.~(\ref{eqosc2a})) modes.
For tensor modes, this method does not yield a new mode (the asymptotic series exhausts the spectrum).
Our analytic expressions were in agreement with numerical results \cite{CKL,bibgub,bib4d}.
They also agreed perfectly with the results from linearized hydrodynamics of the gauge theory fluid on the boundary of AdS space \cite{bibgub,bib4d} in accordance with the AdS/CFT correspondence.

It would be interesting to extend these calculations to a less symmetric configuration that would better fit the experimental setup of heavy ion collisions at RHIC and the LHC.
Then a comparison with experimental results would enhance our understanding of gauge theory fluid dynamics at strong coupling.

\section*{Acknowledgments}

I wish to thank S.~S.~Gubser for useful discussions.

\end{document}